\documentstyle[12pt]{article} 
 
\newtheorem{th}{Theorem}     
\newtheorem{ax}{Axiom}  
\newtheorem{lm}{Lemma} 
\newtheorem{df}{Definition}    
\newtheorem{pr}{Proposition} 
\newtheorem{cl}{Corollary}  
\newtheorem{re}{Remark}    
\newtheorem{as}{Assumption}  
\newtheorem{ex}{Example}
\newcommand{\bth}{\begin{th}\hspace{-5pt}{\bf .} \ } 
\newcommand{\eth}{\end{th}}
\newcommand{\bax}{\begin{ax}\hspace{-5pt}{\bf .} \ } 
\newcommand{\eax}{\end{ax}}
\newcommand{\blm}{\begin{lm}\hspace{-5pt}{\bf .} \ }
\newcommand{\elm}{\end{lm}}
\newcommand{\bdf}{\begin{df}\hspace{-5pt}{\bf .} \ }   
\newcommand{\edf}{\end{df}} 
\newcommand{\bpr}{\begin{pr}\hspace{-5pt}{\bf .} \ } 
\newcommand{\epr}{\end{pr}}
\newcommand{\bcl}{\begin{cl}\hspace{-5pt}{\bf .} \ } 
\newcommand{\ecl}{\end{cl}}
\newcommand{\bre}{\begin{re}\hspace{-5pt}{\bf .} \ }
\newcommand{\ere}{\end{re}}
\newcommand{\bas}{\begin{as}\hspace{-5pt}{\bf .} \ }
\newcommand{\eas}{\end{as}}        
\newcommand{\bex}{\begin{ex}\hspace{-5pt}{\bf .} \ }
\newcommand{\eex}{\end{ex}}
\newcommand{\bpf}{\noindent {\it Proof}\hspace{0.1truecm}: }
\newcommand{\epf}{\hfill${\scriptstyle\diamondsuit\diamondsuit
\diamondsuit}$\par\vspace{1.8mm}\noindent}
\newcommand{\bit}{\begin{itemize}}
\newcommand{\eit}{\end{itemize}\par\noindent}
\newcommand{\beq}{\begin{equation}} 
\newcommand{\eeq}{\end{equation}\par\noindent}
\newcommand{\beqa}{\begin{eqnarray*}}
\newcommand{\eeqa}{\end{eqnarray*}\par\noindent}
\newcommand{\beqn}{\begin{eqnarray}}
\newcommand{\eeqn}{\end{eqnarray}\par\noindent}
\newcommand{\r}{{\cal R}}
\renewcommand{\c}{{\cal C}}
\newcommand{\p}{{\cal P}}
\newcommand{\f}{{\cal F}}

\begin{document}   

\pagenumbering{arabic}   
\pagestyle{plain}


\hbox{}
\par\vskip 2.25 truecm\par      
\noindent{\bf OPERATIONAL RESOLUTIONS AND STATE TRANSI-}  
\par\vskip 0.1 truecm\par
\noindent{\bf TIONS IN A CATEGORICAL
SETTING}\footnote{{\normalsize Camera ready manuscript; In: {\it
Found.~Phys.~Lett.} {\bf 12}, 29--49, 1999.}}
\par
\vspace{0.631cm} 
\par\parindent=1.6cm {\bf Bob Coecke and Isar Stubbe}
\par\vskip 0.406 truecm\par  {\it  FUND-DWIS, Free University of
Brussels,
\par Pleinlaan 2,  B-1050 Brussels, Belgium. 
\par bocoecke@vub.ac.be
\par\vskip 0.406 truecm\par  and   
\par\vskip 0.406 truecm\par  AGEL-MAPA, Universit\'e Catholique
de  Louvain, 
\par Ch$.$ du Cyclotron 2, B-1348 Louvain-La-Neuve, Belgium.  
\par i.stubbe@agel.ucl.ac.be }
\par\vskip 0.406 truecm\par
\par\vskip 0.406 truecm\par  
\par\vskip 0.406 truecm\par  
\indent  Received 1 June 1998; revised 6 December 1998.
\par\vskip 0.406 truecm\par
\par\vskip 0.406 truecm\par
\noindent We define a category with as objects operational
resolutions and with as morphisms --- not necessarily
deterministic --- state transitions. We study connections with
closure spaces and join-complete lattices and sketch physical
applications related to evolution and compoundness.  An appendix
with preliminaries on quantaloids is included.
\par\vskip 0.406 truecm\par
\par\vskip 0.406 truecm\par
\parindent=0.8cm
\noindent   {\bf 1. INTRODUCTION}  
\par\vskip 0.406 truecm\par
\par
\noindent  The core of the mathematical development in this paper
consists of lifting  the --- categorically --- equivalent
descriptions of physical systems by a    (i) `state space' or a 
(ii) `property lattice' --- see [14,20,25,26] --- to an
asymmetrical  --- i.e., not anymore isomorphic --- duality on the
level of: 
\par\smallskip\par
\noindent (i)$^{bis}$ {\it `possible state' transitions ---
`possible' in the sense that an arbitrary initial state is mapped
onto all possible outcome states for this particular `not
necessarily deterministic' transition, i.e., we consider the
propagation of those states that are ``possibly true'' with
respect to the indeterministic nature of this transition},
\par\smallskip\par
\noindent and,   
\par\smallskip\par
\noindent (ii)$^{bis}$ {\it `definite property' transitions ---
`definite' in the sense that we consider the propagation of only
those properties that are ``true with certainty'', even when the
transition is not deterministic.} 
\par\smallskip\par
\noindent However, our mathematical  setup is somewhat more
general than the one in [14,20,25,26] since we consider any set
equipped with a closure operator as a state space and any
complete lattice as a property lattice.  We reach this goal by
demanding that for a `possible state transition', a `definite
property transition' is well-defined and preserves the lattice
join.    It can indeed be physically be motivated that property
transitions should be described by join preserving maps, or
equivalently in the case of complete orthomodular property
lattices, by a complete Baer*-semigroup of hemimorphisms
[3,11,12,13,15,29].  We now briefly sketch a physical
argumentation for this fact along the lines of [15]$^1$.  With an
evolution from time
$t_0$ to time $t_1$ we can associate a map $f^*: {\cal L} \to
{\cal L}$ with
$a_0=f^*(a_1)$ being the cause of $a_1$ in the property lattice
${\cal L}$ on $t_0$, that is, $a_0$ is the weakest property in
${\cal L}$ on time
$t_0$ whose actuality --- read `being true' --- guarantees the
actuality of $a_1$ on time
$t_1$.  The fact that the lattice meet is nothing else than the
semantic   conjunction [27] then implies that $f^*$ preserves
non-empty meets.  As a consequence, it has a join preserving
Galois dual [4,17,21]: 
\beq f:[0,f^*(1)]\to{\cal L}:a_0\mapsto\wedge\{a_1\in{\cal L}\mid
a_0\leq f^*(a_1)\}
\eeq  that exactly expresses the propagation of the properties:
$f$ maps an arbitrary property $a_0$ to the strongest one ---
expressed as a meet --- whose actuality is implied by the
actuality of $a_0$.  Conversely, to any such join preserving map
$f$, expressing propagation, we can associate a meet preserving
map $f^*$ that expresses the physically justifiable assignation
of temporal causes for that evolution.  In [15] however, only so
called ``strong deterministic evolutions'' that send atoms to
atoms have been considered, as such excluding the description of
indeterministic state transitions.   In our approach we formally
integrate indeterministic transitions by considering maps on
powersets of a state space, as it will be discussed in more
detail in the 5th section of this paper.   Besides this
indeterministic aspect we will also consider state transitions
with non-equal domain and codomain for essentially two reasons:
(i) some externally induced state transitions might correspond
with an actual change of the state space of the system; (ii) maps
between different property lattices provide an appropriate
structure for the description of mutually induced state
transitions between individual entities within a compound system
--- a proof for the existence of such a representation for
compound quantum systems can be found in [7,9].  For more details
on this aspect we refer to the 6th section of this paper.  The
main motivation for a categorical  treatment of state transitions
is the observation that  they compose in a natural way by
consecutive application,  and that composition of morphisms is
exactly the structural ingredient that constitutes a category
[1,5,19].   As such, it is in our particular setup very natural
to express  state transitions as morphisms of a category, where
the objects describe the states and properties of a physical
system. Functorality of maps between categories then expresses
preservation of  consecutive application, in our case by coupling
`possible state transitions' and corresponding `definite property
transitions'.
\par\vskip 0.406 truecm\par
\par\vskip 0.406 truecm\par
\noindent  {\bf 2. OPERATIONAL RESOLUTIONS}
\par\vskip 0.406 truecm\par
\par 
\noindent    We begin by defining the objects of our categories. 
\bdf\label{opres}    For a given set $\Sigma$, an `operational
resolution' is defined as a map ${\cal C}_{pr}:{\cal
P}(\Sigma)\rightarrow {\cal L}$ with as domain the powerset of
$\Sigma$ and as codomain a poclass$^2$
$({\cal L},\leq)$, such that for all $T,T',T_i\in{\cal
P}(\Sigma)$:
\beqn\label{df:opres2} T\subseteq T'\ \ \ \ \ &\Rightarrow&\ {\cal
C}_{pr}(T)\leq{\cal C}_{pr}(T')\\
\label{df:opres3}
\forall i:{\cal C}_{pr}(T_i)\leq{\cal C}_{pr}(T)&\Rightarrow&{\cal
C}_{pr}(\cup_iT_i)\leq{\cal C}_{pr}(T)\\
\label{df:opres3'} T\not=\emptyset\ \ \ \ \ \ &\Rightarrow&\
\ {\cal C}_{pr}(T)\not={\cal C}_{pr}(\emptyset)  
\eeqn
\edf      The chosen axioms can be verified physically by
interpreting ${\cal C}_{pr}$ as assigning to a collection of
`possible states'
$T\in{\cal P}(\Sigma)$ the  smallest --- i.e., strongest ---
`definite property' ${\cal C}_{pr}(T)\in{\cal L}$ physically
implied by every state $p\in T$, i.e, referring to the
terminology of [14,20,25,26], it is the conjunction of all
properties that are always actual whenever at least one  $p\in T$
is actual --- the existence of such a conjunction follows from
Theorem \ref{factorization}.    Let us first recall some basic
results on such operational resolutions [11]. The image of
${\cal C}_{pr}$, which is a subset of the class ${\cal L}$ and
thus inherits the partial order $\leq$, is a complete lattice
with, for any 
$\{T_i\}_i\subseteq{\cal P}(\Sigma)$, $\vee_i{\cal
C}_{pr}(T_i)={\cal C}_{pr}(\cup_iT_i)$, bottom element
${\cal C}_{pr}(\emptyset)$ and top element ${\cal
C}_{pr}(\Sigma)$. Also,
$\{{\cal C}_{pr}(p)\mid p\in\Sigma\}$ is an order generating set
of
$im({\cal C}_{pr})$, in the sense that
$\forall T\in{\cal P}(\Sigma): {\cal C}_{pr}(T)=\vee_{t\in T}{\cal
C}_{pr}(t)$.  Given a set $X$, an operator ${\cal C}:{\cal
P}(X)\rightarrow{\cal P}(X)$ on the powerset of
$X$ is called `closure operator on $X$' if the following are met:
(C1):
$T\subseteq{\cal C}(T)$; (C2):
$T\subseteq T'\Rightarrow {\cal C}(T)\subseteq{\cal C}(T')$; (C3):
${\cal C}({\cal C}(T))={\cal C}(T)$. A closure is called
$T_0$ if moreover ${\cal C}(\emptyset)=\emptyset$ and ${\cal
C}(\{x\})={\cal C}(\{y\})\Rightarrow x=y$ for any $x,y\in X$. It
is called
$T_1$ if ${\cal C}(\emptyset)=\emptyset$ and ${\cal
C}(\{x\})=\{x\}$ for any
$x\in X$. We have that every operational resolution
${\cal C}_{pr}:{\cal P}(\Sigma)\rightarrow{\cal L}$ factors into
a closure operator
${\cal C}:{\cal P}(\Sigma)\rightarrow{\cal
F}(\Sigma)\subseteq{\cal P}(\Sigma)$ with ${\cal
C}(\emptyset)=\emptyset$, and an embedding of the   poset of
${\cal C}$-closed subsets ${\cal F}(\Sigma)$ into the poclass
${\cal L}$, \ 
$\theta:{\cal F}(\Sigma)\rightarrow{\cal L}$, such that
${\cal F}(\Sigma)\cong im(\theta)=im({\cal C}_{pr})$ as lattices.
The closure factor is given by ${\cal C}(T)=\cup\{S\in{\cal
P}(\Sigma)\mid{\cal C}_{pr}(S)={\cal C}_{pr}(T)\}$, which can be
rewritten as
${\cal C}(T)=\{t\in\Sigma\mid{\cal C}_{pr}(t)\leq{\cal
C}_{pr}(T)\}$. The prescription of the embedding
$\theta$ is
$\theta(F)={\cal C}_{pr}(F)$, for any $F\in{\cal F}(\Sigma)$. It
can be verified that this factorization is unique. Conversely to
the factorization, any closure space
$(\Sigma,{\cal C})$ for which ${\cal C}(\emptyset)=\emptyset$ and
any embedding of the poset of ${\cal C}$-closed subsets of
$\Sigma$ in a poclass ${\cal L}$, say \ $\theta:{\cal
F}(\Sigma)\rightarrow{\cal L}$, such that ${\cal F}(\Sigma)\cong
im(\theta)$ as lattices, uniquely define an operational
resolution, namely
${\cal C}_{pr}=\theta\circ{\cal C}$. Referring to these results,
we state the following theorem.
\bth\label{factorization}   Given a set $\Sigma$ and a poclass
${\cal L}$, a map
${\cal C}_{pr}:{\cal P}(\Sigma)\rightarrow{\cal L}$ is an
operational resolution if and only if it factors uniquely into a
closure
${\cal C}$ with
${\cal C}(\emptyset)=\emptyset$, and an embedding
$\theta$ of the poset of \ ${\cal C}$-closed subsets
${\cal F}(\Sigma)$ into the poclass ${\cal L}$ such that the
image of
$\theta$, inheriting the order from
${\cal L}$, is a complete lattice that is isomorphic to
${\cal F}(\Sigma)$.
\eth  This means that
${\cal C}_{pr}:{\cal P}(\Sigma)\rightarrow{\cal L}$ can be
characterized either by the triple $(\Sigma, {\cal L}, {\cal
C}_{pr})$ or by the quadruple
$(\Sigma, {\cal L}, {\cal C},
\theta)$.

Next we want to elaborate on ``how an operational resolution
orders and separates points'', much in analogy to closure
operators. These considerations will lead to the notion of a
`canonical resolution'.
\bdf\label{T} An operational resolution
${\cal C}_{pr}:{\cal P}(\Sigma)\rightarrow{\cal L}$ is a
$T_0$-resolution if \ 
$\forall p,q\in\Sigma: {\cal C}_{pr}(p)={\cal
C}_{pr}(q)\Rightarrow p= q$; it is a
$T_1$-resolution if \ 
$\forall p,q\in\Sigma: {\cal C}_{pr}(p)\leq{\cal
C}_{pr}(q)\Rightarrow p=q$.
\edf It can be verified straightforwardly, and it justifies the
terminology in Definition \ref{T}, that
${\cal C}_{pr}$ is $T_0$ ($T_1$) if and only if its closure factor
${\cal C}$ is $T_0$ ($T_1$). Concerning an operational resolution
${\cal C}_{pr}:{\cal P}(\Sigma)\rightarrow{\cal L}$, we further
introduce the following notations, for $p,q\in\Sigma$:
\[ 
\left\{\begin{array}{ccc}  p\lhd_{pr}q & \Leftrightarrow & {\cal
C}_{pr}(p)<{\cal C}_{pr}(q)\hfill 
\\  p=_{pr}q & \Leftrightarrow & {\cal C}_{pr}(p)={\cal
C}_{pr}(q)\hfill\\  p\unlhd_{pr}q & \Leftrightarrow & p\lhd_{pr}q
\ \ or \ \  p=_{pr}q
\end{array}
\right.
\]   This defines a preordered set
$(\Sigma,\unlhd_{pr})$. It obviously follows that
${\cal C}_{pr}$ is $T_0$ if and only if
$[p=_{pr}q\Rightarrow p=q]$ for all $p,q\in\Sigma$ and that
${\cal C}_{pr}$ is $T_1$ if and only if
$[p\unlhd_{pr}q\Rightarrow p=q]$ for all
$p,q\in\Sigma$. The following examples prove their importance
further in this text.
\bex\label{canonicalexample} If
$\Sigma$ is a `full set of states' $[2,26]$\ for a complete
lattice
${\cal L}$ --- i.e., $\Sigma$ is a subset of ${\cal L}$, not
containing the bottom element, such that
\ $t=\vee\{a\in\Sigma\mid a\leq t\}$ for all $t\in {\cal L}$ ---
then
${\cal C}_{pr}:{\cal P}(\Sigma)\rightarrow {\cal L}:T\mapsto\vee
T$ is an operational resolution.
$\Sigma$ inherits order from ${\cal L}$, and this order coincides
with
$\unlhd_{pr}$ (from a slightly different viewpoint we could also
say that this operational resolution ``recuperates'' the {\it a
priori} order on
$\Sigma$, which is inherited from ${\cal L}$ through
set-inclusion). This operational resolution is always $T_0$. If
${\cal L}$ is atomistic and $\Sigma$ is its set of atoms, then
and only then it is $T_1$.
\eex   This example exhibits how the notion of ``operational
resolution'' generalizes the state/property duality as it is put
forward in [20].
\bex\label{closuresasresolutions}  Given any closure ${\cal C}$
on a set
$\Sigma$ such that ${\cal C}(\emptyset)=\emptyset$, ${\cal
C}:{\cal P}(\Sigma)\to{\cal F}(\Sigma)$ --- where ${\cal
F}(\Sigma)$ is the family of ${\cal C}$-closed subsets of
$\Sigma$ --- defines an operational resolution. This operational
resolution is obviously
$T_0$ $(T_1)$ whenever ${\cal C}$ is so as closure.   
\eex  By definition of
$\unlhd_{pr}$, the surjection
$\Sigma\rightarrow\{{\cal C}_{pr}(p)\mid
p\in\Sigma\}:p\mapsto{\cal C}_{pr}(p)$ preserves
$\unlhd_{pr}$. This map is injective, thus bijective exactly for
$T_0$ resolutions. We have that
$\{{\cal C}_{pr}(p)\mid p\in\Sigma\}\subseteq im({\cal
C}_{pr})\setminus\{{\cal C}_{pr}(\emptyset)\}$, but if this
inclusion 'saturates' to an equality, then we have a very
particular kind of operational resolution at hand.
\bdf 
${\cal C}_{pr}:{\cal P}(\Sigma)\rightarrow {\cal L}$ will be
called `saturated' if the map $\Sigma\rightarrow im({\cal
C}_{pr})\setminus\{{\cal C}_{pr}(\emptyset)\}:p\mapsto{\cal
C}_{pr}(p)$ is surjective. An operational resolution that is
saturated and
$T_0$, will be called `canonical'.
\edf     If ${\cal C}_{pr}$ is canonical then, and only then, we
have isomorphic lattices
$(\Sigma\cup\{0\},\unlhd_{pr})\cong (im({\cal C}_{pr}),\leq)$
where we define that $0\unlhd_{pr}p$ for all $p\in\Sigma$, and
where
$\leq$ on $im({\cal C}_{pr})$ is inherited from ${\cal L}$. 
\bex\label{canonicalexample2}  The operational resolution of
Example
\ref{canonicalexample} is canonical if and only if
$\Sigma={\cal L}\setminus\{0\}$.  
\eex  For any operational resolution ${\cal C}_{pr}:{\cal
P}(\Sigma)\rightarrow{\cal L}$, we can define a canonical
resolution ${\cal C}'_{pr}:{\cal P}(\Sigma')\rightarrow{\cal L'}$
such that there exists a map
$\phi:\Sigma\rightarrow\Sigma'$ fulfilling
$\forall T\in{\cal P}(\Sigma)$: ${\cal C}_{pr}(T)={\cal
C}'_{pr}(\{\phi(t)\mid t\in T\})$, that is:
\[
\begin{array}{cccccccc}
 & \ \ \ \Sigma & & & {\cal P}(\Sigma) & 
\hspace{-1mm}\stackrel{{\cal C}_{pr}}{\longrightarrow}\  &
im({\cal C}_{pr}) & \\ exists\  & \phi\downarrow &\ such\
that\hspace{-1mm} & &  {\cal P}(\phi)\downarrow\ \ \ \  & & || &
commutes, \\
 & \ \ \ \Sigma' & & & {\cal P}(\Sigma') & 
\hspace{-1mm}\stackrel{{\cal C}'_{pr}}{\longrightarrow}\   &
im({\cal C}'_{pr}) & \\
\end{array}    
\]    where ${\cal P}(\phi)(T)=\{\phi(t)\mid t\in T\}$. Indeed,
an obvious example of such a construction is the following:\[
\left\{
\begin{array}{c}
\Sigma'=im({\cal C}_{pr})\setminus\{{\cal C}_{pr}(\emptyset)\} \\ 
{\cal L}'=im({\cal C}_{pr})\hfill 
\end{array}
\right.
\ \ \ \ \   
\left\{
\begin{array}{c}
\phi:\Sigma\rightarrow\Sigma':t\rightarrow{\cal
C}_{pr}(\{t\})\hfill \\  {\cal C}'_{pr}:{\cal
P}(\Sigma')\rightarrow{\cal L}':T\mapsto\vee T
\end{array}
\right.  
\]    Moreover, such a canonical resolution is determined up to an
isomorphism of its domain and a choice for
${\cal L}'$. Indeed, let ${\cal C}''_{pr}:{\cal
P}(\Sigma'')\rightarrow{\cal L}''$ be another canonical
resolution determined by ${\cal C}_{pr}:{\cal
P}(\Sigma)\rightarrow{\cal L}$, then by definition
$\Sigma''\cup\{0''\}\cong im({\cal C}''_{pr})=im({\cal
C}_{pr})=im({\cal C}'_{pr})\cong\Sigma'\cup\{0'\}$, thus there is
a bijection
$\xi:\Sigma'\rightarrow\Sigma''$ which implies that there is an
isomorphism of lattices ${\cal P}(\xi):{\cal
P}(\Sigma')\stackrel{\sim}{\longrightarrow}{\cal P}(\Sigma'')$. We
formulate the net result of the above reasoning as a theorem.
\bth\label{resolutiondefinescanonical} Every operational
resolution defines an essentially \\ unique canonical one.
\eth
\par\vskip 0.406 truecm\par
\par\vskip 0.406 truecm\par
\noindent   {\bf 3. STATE TRANSITIONS AS MORPHISMS}
\par\vskip 0.406 truecm\par
\par 
\noindent  Consider the collection of all $(\Sigma,{\cal L},{\cal
C}_{pr})$   such that 
${\cal C}_{pr}:{\cal P}(\Sigma)\rightarrow{\cal L}$ is an
operational resolution. This will be the object collection of a
first category. To deal with the morphisms between two such
triples $(\Sigma_1,{\cal L}_1,{\cal C}_{pr,1})$ and
$(\Sigma_2,{\cal L}_2,{\cal C}_{pr,2})$ we first introduce the
following notations, applying on maps 
$f:{\cal P}(\Sigma_1)\rightarrow{\cal P}(\Sigma_2)$:
\par\medskip\noindent
$A_{\cup}$: $\forall\{T_i\}_i\subseteq{\cal P}(\Sigma_1):
f(\cup_iT_i)=\cup_if(T_i)$;
\par\noindent
$A_{\emptyset}$: $\forall T\in{\cal P}(\Sigma_1):   
f(T)=\emptyset\Leftrightarrow T=\emptyset$;
\par\noindent
$A_{\#}$: $\forall T,T'\in{\cal P}(\Sigma_1): {\cal
C}_{pr,1}(T)={\cal C}_{pr,1}(T')\Rightarrow{\cal
C}_{pr,2}(f(T))={\cal C}_{pr,2}(f(T'))$. 
\par\medskip\noindent   Mathematically, these three conditions
encode the ``structure preserving'' nature of a map $f:{\cal
P}(\Sigma_1)\rightarrow{\cal P}(\Sigma_2)$ with respect to the
objects `operational resolutions' expressed as triples
$(\Sigma,{\cal L},{\cal C}_{pr})$, in a way that will become
clear in Proposition \ref{Fpr}. 
\bpr\label{propquantaloid}   We can define a quantaloid
$\underline{Res}^{\#}_{\emptyset}$, in which the join of maps is
computed pointwise, by taking as object class the collection of
triples
$(\Sigma,{\cal L},{\cal C}_{pr})$ such that ${\cal C}_{pr}:{\cal
P}(\Sigma)\rightarrow{\cal L}$ is an operational resolution, and
taking as hom-set between any two such objects $(\Sigma_1,{\cal
L}_1,{\cal C}_{pr,1})$ and $(\Sigma_2,{\cal L}_2,{\cal
C}_{pr,2})$:
$$\{f:{\cal P}(\Sigma_1)\rightarrow{\cal P}(\Sigma_2)\mid
im(f)=\emptyset\ or\  f\ meets\ A_{\cup},A_{\emptyset},A_{\#}\}
$$
\epr
\bpf  (o) In this proof, as in all others, the cases where the
``bottom'' morphism, given by the underlying map
$\emptyset:{\cal P}(\Sigma_1)\rightarrow{\cal
P}(\Sigma_2):T\mapsto\emptyset$, comes into play are trivial, so
we do not consider them. But it is important to include this map
in each hom-set for these to have a bottom element. (i) Identity
morphisms --- morphisms of which the underlying map is the
identity --- obviously meet all three conditions, and for
composable morphisms
$f_1,f_2$ --- morphisms with composable underlying maps --- the
composite meets all conditions:
$(f_2\circ f_1)(\cup_iT_i) =f_2(\cup_i(f_1(T_1)))
=\cup_if_2(f_1(T_i) =\cup_i(f_2\circ f_1)(T_i)$, also $ (f_2\circ
f_1)(T)=\emptyset
\Leftrightarrow f_1(T)=\emptyset
\Leftrightarrow T=\emptyset
$ and finally ${\cal C}_{pr,1}(T)={\cal C}_{pr,1}(T')
\Rightarrow {\cal C}_{pr,2}(f_1(T))={\cal C}_{pr,2}(f_1(T'))
\Rightarrow{\cal C}_{pr,3}((f_2\circ f_1)(T))={\cal
C}_{pr,3}((f_2\circ f_1)(T'))$. (ii) Pointwise joins of maps
exist: for morphisms
$f_i:(\Sigma_1,{\cal L}_1,{\cal C}_{pr,1})\rightarrow
(\Sigma_2,{\cal L}_2,{\cal C}_{pr,2})$ we have: $
(\bigvee_if_i)(\cup_jT_j)=
\cup_{i,j}f_i(T_j)=
\cup_j(\bigvee_if_i)(T_j)$ and
$(\bigvee_if_i)(T)=\emptyset\Leftrightarrow\cup_if_i(T)
=\emptyset\Leftrightarrow\forall i:f_i(T)=\emptyset\Leftrightarrow
T=\emptyset$ and finally also 
${\cal C}_{pr,1}(T)={\cal C}_{pr,1}(T')
\Rightarrow\forall i:{\cal C}_{pr,2}(f_i(T))={\cal
C}_{pr,2}(f_i(T'))$ which implies $ {\cal
C}_{pr,2}((\bigvee_if_i)(T)) ={\cal C}_{pr,2}(\cup_if_i(T))
=\vee_i{\cal C}_{pr,2}(f_i(T)) =\vee_i{\cal C}_{pr,2}(f_i(T'))
={\cal C}_{pr,2}(\cup_if_i(T')) ={\cal
C}_{pr,2}((\bigvee_if_i)(T'))$. (iii) Consider a collection of
'parallel' morphisms
$f_i:(\Sigma_1,{\cal L}_1,{\cal C}_{pr,1})\rightarrow
(\Sigma_2,{\cal L}_2,{\cal C}_{pr,2})$ and a composable morphism
$g:(\Sigma_2,{\cal L}_2,{\cal C}_{pr,2})\rightarrow(\Sigma_3,{\cal
L}_3,{\cal C}_{pr,3})$, then we have that:
$(g\circ(\bigvee_if_i))(-)=g(\cup_i(f_i(-)))=\cup_i(g(f_i(-)))=
\bigvee_i(g\circ f_i)(-)$. Likewise for the distributivity on the
right.
\epf     First note that we had to ``add'' a bottom element to
our collection of maps in order to obtain a complete lattice ---
this bottom element is the empty union of maps. When we interpret
the above maps as state transitions, something that will be
discussed in detail bellow, this bottom element in a lattice of
state transitions ``plays the same role'' as the bottom  element
in a property lattice: the latter stands for the ``absurd
property'' which a system will never have; as such can the zero
map be interpreted as an ``absurd transition''. In the next
section we will reconsider this aspect and show that it makes
sense to introduce ``partially absurd transitions''.  

The very idea behind ``operational resolution'' --- assigning to
any subset
$T$ of a system's state set $\Sigma$ a strongest property of that
system implied by all the states in $T$ --- suggests that any
morphism
$f\in\underline{Res}_{\emptyset}^{\#}((\Sigma_1,{\cal L}_1,{\cal
C}_{pr,1}),(\Sigma_2,{\cal L}_2,{\cal C}_{pr,2}))$ is
``reflected'' through the given operational resolutions as:
$$f_{pr}:im({\cal C}_{pr,1})\rightarrow im({\cal C}_{pr,2}):{\cal
C}_{pr,1}(T)\mapsto{\cal C}_{pr,2}(f(T)),$$   yielding exactly
the corresponding `definite property transition'. Indeed, if the
strongest --- definite --- actual property of a system, initially
in a state that is contained in a
$T\subseteq\Sigma$, is
${\cal C}_{pr,1}(T)\in im({\cal C}_{pr,1})$, then after the
``change of state''
$f$ the state of the system is in $f(T)$, thus with  strongest
--- definite --- actual property ${\cal C}_{pr,2}(f(T))$. It is
then exactly condition $A_{\#}$ on the morphism $f$ that assures
us that $f_{pr}$ is well-defined: for $a\in im({\cal C}_{pr,1})$
the value of
$f_{pr}(a)$ does not depend on the ``representative''
$T\in\Sigma_1$ that we choose such that ${\cal C}_{pr,1}(T)=a$.
In other terms, it implies that the following diagram commutes:
\[
\begin{array}{ccc} {\cal P}(\Sigma_1) &
\stackrel{f}{\longrightarrow} & {\cal P}(\Sigma_2) \\ {\cal
C}_{pr,1}
\downarrow & & \downarrow {\cal C}_{pr,2} \\ im({\cal C}_{pr,1}) &
\stackrel{f_{pr}}{\longrightarrow} & im({\cal C}_{pr,2}) \\
\end{array}  
\]  Further it can be verified that such an $f_{pr}$, which is a
map between join complete lattices, preserves joins: 
$f_{pr}(\vee_i{\cal C}_{pr,1}(T_i))=f_{pr}({\cal
C}_{pr,1}(\cup_iT_i))={\cal C}_{pr,2}(f(\cup_iT_i))={\cal
C}_{pr,2}(\cup_if(T_i))=\vee_i{\cal
C}_{pr,2}(f(T_i))=\vee_if_{pr}({\cal C}_{pr,1}(T_i))$. Taking
into account the arguments in the introduction on the propagation
of properties we can interpret these formal results.
\par
\bigskip
\par\noindent {\it 
$\underline{Physically\ conclusive}$:  Conditions $A_\#$ and
$A_\cup$ assure that a `possible state transition'
$f:{\cal P}(\Sigma_1)\rightarrow{\cal P}(\Sigma_2)$ determines a
unique join preserving `definite property transition'
$f_{pr}:im({\cal C}_{pr,1})\to im({\cal C}_{pr,2})$.}
\par
\bigskip
\par\noindent Note that $f_{pr}$ maps the bottom of its domain
exactly onto the bottom of its codomain:
$f_{pr}({\cal C}_{pr,1}(T))=0_2\Leftrightarrow {\cal
C}_{pr,2}(f(T))=0_2\Leftrightarrow f(T)=\emptyset\Leftrightarrow
T=\emptyset\Leftrightarrow {\cal C}_{pr,1}(T)=0_1$. We give these
conditions applying on a map $g:im({\cal C}_{pr,1})\rightarrow
im({\cal C}_{pr,2})$ a notation:
\par\medskip\noindent
$A_{\vee}$: $\forall\{a_i\}_i\subseteq im({\cal C}_{pr,1}):
g(\vee_ia_i)=\vee_ig(a_i)$;
\par\noindent
$A_0$: $\forall a\in im({\cal C}_{pr,1}): g(a)=0_1\Leftrightarrow
a=0_2$.  
\bpr\label{Fpr}  We can define a quantaloid
$\underline{Res}_0$, in which the join of maps is computed
pointwise, by taking as object class the collection of triples
$(\Sigma,{\cal L},{\cal C}_{pr})$ such that ${\cal C}_{pr}:{\cal
P}(\Sigma)\rightarrow{\cal L}$ is an operational resolution, and
taking as hom-set between any two such objects $(\Sigma_1,{\cal
L}_1,{\cal C}_{pr,1})$ and $(\Sigma_2,{\cal L}_2,{\cal
C}_{pr,2})$:
$$\{f:im({\cal C}_{pr,1})\rightarrow im({\cal C}_{pr,2})\mid
im(f)=\{0\}\  or\ f\ meets\ A_{\vee},A_0\}$$ And the following
action on an object $(\Sigma,{\cal L},{\cal C}_{pr})$ and a
morphism $f:(\Sigma_1,{\cal L}_1,{\cal
C}_{pr,1})\to(\Sigma_2,{\cal L}_2,{\cal C}_{pr,2})$ in
$\underline{Res}_0$ defines a full bijective quantaloid morphism 
$F_{pr}:\underline{Res}_{\emptyset}^{\#}\rightarrow\underline{Res}_0$:
\[  
\left\{\begin{array}{l} F_{pr}(\Sigma,{\cal L},{\cal
C}_{pr})=(\Sigma,{\cal L},{\cal C}_{pr})\\ 
F_{pr}(f)=f_{pr}:im({\cal C}_{pr,1})\rightarrow im({\cal
C}_{pr,2}):{\cal C}_{pr,1}(T)\mapsto{\cal C}_{pr,2}(f(T))
\end{array}
\right.
\]
\epr
\bpf   We leave the straightforward verification that
$\underline{Res}_0$ is a quantaloid to the reader. The action of
$F_{pr}$ on objects is simply the identity, so nothing to verify
there. The above remarks point out that the action on morphisms
is well-defined, and that indeed any
$f_{pr}$ is a morphism of $\underline{Res}_0$. We now prove
functorality: Since the underlying map of an identity morphism is
an identity, it follows that
$F_{pr}$ preserves identities, and pasting together commutative
diagrams:
\[
\begin{array}{ccccc} {\cal P}(\Sigma_1) &
\stackrel{f_1}{\longrightarrow} & {\cal P}(\Sigma_2) &
\stackrel{f_2}{\longrightarrow} & {\cal P}(\Sigma_3) \\ {\cal
C}_{pr,1} 
\downarrow & & \downarrow {\cal C}_{pr,2} & & \downarrow {\cal
C}_{pr,3} \\ im({\cal C}_{pr,1}) &
\stackrel{f_{1,pr}}{\longrightarrow} & im({\cal C}_{pr,2}) &
\stackrel{f_{2,pr}}{\longrightarrow} & im({\cal C}_{pr,3}) \\
\end{array}  
\]  yields $F_{pr}(f_2\circ f_1)=F_{pr}(f_2)\circ F_{pr}(f_1)$.
$F_{pr}$ induces $\bigvee$-preserving maps on hom-sets: consider
$f_i:(\Sigma_1,{\cal L}_1,{\cal
C}_{pr,1})\rightarrow(\Sigma_2,{\cal L}_2,{\cal C}_{pr,2})$ in
$\underline{Res}_{\emptyset}^{\#}$, then
$(\bigvee_if_i)_{pr}({\cal C}_{pr,1}(T))={\cal
C}_{pr,2}(\cup_if_i(T))=\vee_i{\cal
C}_{pr,2}(f_i(T))=\vee_i(f_{i,pr}({\cal
C}_{pr,1}(T)))\\=(\bigvee_if_{i,pr})({\cal C}_{pr,1}(T))$.
Further consider the following situation, where $g$ is a given
$\underline{Res}_0$-morphism:
\[
\begin{array}{ccc} {\cal P}(\Sigma_1) &
\stackrel{\exists ? g^*}{\longrightarrow} & {\cal P}(\Sigma_2)
\\  {\cal C}_{pr,1}\downarrow \
\ \ \ \ \ & & \ \ \ \ \downarrow {\cal C}_{pr,2} \\ im({\cal
C}_{pr,1}) &
\stackrel{g}{\longrightarrow} & im({\cal C}_{pr,2}) \\
\end{array}  
\]     Remembering the factorization of an operational
resolution, {\it in casu} ${\cal C}_{pr,2}=\theta_2\circ{\cal
C}_2$, it makes sense to  define: 
\beqa g^*:{\cal P}(\Sigma_1)\rightarrow{\cal
P}(\Sigma_2):T\mapsto\cup_{t\in T}(\theta_2^{-1}\circ g\circ{\cal
C}_{pr,1})(t)
\eeqa and thus $g^*$ is (the underlying map of) a morphism from
$(\Sigma_1,{\cal L}_1,{\cal C}_{pr,1})$ to
$(\Sigma_2,{\cal L}_2,{\cal C}_{pr,2})$ in
$\underline{Res}_{\emptyset}^{\#}$:
$g^*(\cup_iT_i)=\cup_ig(T_i)$ is obvious;
$g^*(T)=\emptyset\Leftrightarrow\forall t\in T:(\theta_2^{-1}\circ
g\circ{\cal C}_{pr,1})(t)=\emptyset\Leftrightarrow\forall t\in
T:g({\cal C}_{pr,1}(t))=0_2\Leftrightarrow\forall t\in T:{\cal
C}_{pr,2}(t)=0_1\Leftrightarrow T=\emptyset$; the following
square commutes:
\[
\begin{array}{ccc}    {\cal P}(\Sigma_1) & 
\stackrel{g^*}{\longrightarrow} & {\cal P}(\Sigma_2) \\ {\cal
C}_{pr,1} \downarrow & & \downarrow {\cal C}_{pr,2} \\  im({\cal
C}_{pr,2}) & \stackrel{g}{\longrightarrow} & im({\cal C}_{pr,1})
\\
\end{array}  
\]   since ${\cal C}_{pr,2}(g^*(t))=({\cal
C}_{pr,2}\circ\theta_2^{-1}\circ g\circ{\cal
C}_{pr,1})(t)=g({\cal C}_{pr,1}(t))$ for any $t\in\Sigma_1$,
which implies that for any $T\subseteq\Sigma_1$ also ${\cal
C}_{pr,2}(g^*(T))=g({\cal C}_{pr,1}(T))$. This proves at once
$A_{\#}$ and
$F_{pr}(g^*)=g$. Thus
$F_{pr}$ restricted to hom-sets is surjective.
\epf
\par
\medskip
\par\noindent {\it 
$\underline{Physically\ conclusive}$:  `possible state
transitions' and implied `definite property transitions' are in
categorical correspondence under the binary operation
`composition of maps' that formally implements consecution of
transitions.}
\par
\bigskip
\par\noindent We will now comment on the construction in the
proof of the ``reciprocal''
$g^*$ for a given
$g$ in $\underline{Res}_0$. Since the restriction of
$F_{pr}$ to hom-sets:
\[
\begin{array}{rl} F_{pr} & :\underline{Res}_{\emptyset}^{\#}(-,-)
\rightarrow\underline{Res}_0(F_{pr}(-),F_{pr}(-))
\end{array}
\]   is join-preserving, it has a unique meet-preserving Galois
dual [4,17,21]:
\[
\begin{array}{rl}   F_{pr}^* & 
:\underline{Res}_0(F_{pr}(-),F_{pr}(-))
\rightarrow
\underline{Res}_{\emptyset}^{\#}(-,-):\vspace{2mm}\\
 & 
\ \ \ \ \  \ \ \
\ \ \ \ \ \ \ \ \ \ \ \ \ \ \ \ \ \ \ \ \
g\mapsto\bigvee\{f\in\underline{Res}_{\emptyset}^{\#}(-,-)\mid
F_{pr}(f)\leq g\}
\end{array}
\]   We can show the following.
\bre  Referring to the above notations we have: 
\par\noindent (i) $g^*=F_{pr}^*(g)$ for any
$g\in\underline{Res}_0(F_{pr}(-),F_{pr}(-))$;
\par\noindent (ii)
$F_{pr}\circ
F_{pr}^*=id:{\underline{Res}_{\emptyset}^{\#}(-,-)}\to
\underline{Res}_{\emptyset}^{\#}(-,-):f\mapsto f$;
\par\noindent (iii) $F_{pr}^*$ preserves composition;
\par\noindent (iv) in general $F_{pr}^*$ is not functoral;  
\par\noindent (v) in general $F_{pr}^*$ does not preserve
arbitrary joins. 
\ere
\bpf  (i) Since
$g^*\in\{f\in\underline{Res}_{\emptyset}^{\#}(-,-)\mid
F_{pr}(f)\leq g\}$, $g^*\leq F_{pr}^*(g)$ is obvious. For any
$f\in\underline{Res}_{\emptyset}^{\#}(-,-)$ such that
$F_{pr}(f)\leq g$ and any $t\in\Sigma_1$ we have that
$f(t)\subseteq(\theta^{-1}_2\circ{\cal C}_{pr,2}\circ
f)(t)\subseteq (\theta_2^{-1}\circ g\circ{\cal
C}_{pr,1})(t)=g^*(t)$, therefore, for any
$T\subseteq\Sigma_1$, also $f(T)=\cup_{t\in T}f(t)
\subseteq\cup_{t\in T}g^*(t)=g^*(T)$. In other terms, $f\leq g^*$
for any such
$f$, thus $F_{pr}^*(g)\leq g^*$. (ii) Corollary of (i). (iii)
Setting that
$F_{pr}^*(g_2\circ g_1)=F_{pr}^*(g_2)\circ F_{pr}^*(g_1)$ we see
that
$F_{pr}(F_{pr}^*(g_2)\circ
F_{pr}^*(g_1))=F_{pr}(F_{pr}^*(g_2))\circ
F_{pr}(F_{pr}^*(g_1))=g_2\circ g_1$. (iv) for $id:im({\cal
C}_{pr})\rightarrow im({\cal C}_{pr})$, $F_{pr}^*(id):{\cal
P}(\Sigma)\rightarrow{\cal P}(\Sigma):T\mapsto\cup\{{\cal
C}(t)\mid t\in T\}$, which in general contains
$T$ but is not necessarily contained in $T$ --- so identities are
not preserved. (v) All we know is that $F_{pr}^*$ preserves
meets, nothing more.
\epf  Conclusion: this Galois dual --- in fact, this right
inverse --- to
$F_{pr}$ cannot be extended to a functor, let alone a quantaloid
morphism. However, from Proposition
\ref{Fpr} we can derive the following equivalence, to be
understood as the categorization of Theorem
\ref{resolutiondefinescanonical}.
\bpr\label{U}  Set that \
$U:\underline{Res}_0\rightarrow\underline{JCLat}_0$ works on
$(\Sigma,{\cal L},{\cal C}_{pr})$ and 
$f\in \underline{Res}_0((\Sigma_1,{\cal L}_1,{\cal
C}_{pr,1}),(\Sigma_2,{\cal L}_2,{\cal C}_{pr,2}))$ respectively
as:
\[
\left\{\begin{array}{l} U(\Sigma,{\cal L},{\cal C}_{pr})=im({\cal
C}_{pr})\\ U(f)=f:im({\cal C}_{pr,1})\rightarrow im({\cal
C}_{pr,2}):{\cal C}_{pr,1}(T)\mapsto{\cal C}_{pr,2}(f(T))
\end{array}
\right.
\]
\noindent This defines a fully faithful surjective quantaloid
morphism.
\epr 
\bpf 
$U$ is surjective on objects because for any given ${\cal L}$,
object of
$\underline{JCLat}_0$, ${\cal C}_{pr}:{\cal P}({\cal
L}\setminus\{0\})\rightarrow {\cal L}:T\mapsto\vee T$, cfr.
Example
\ref{canonicalexample2}, has as image through $U$ exactly
${\cal L}$. The rest is trivial.
\epf  
\bcl\label{CL111} The quantaloids $\underline{Res}_0$ and
$\underline{JCLat}_0$ are equivalent. Moreover, $(U^*\circ
U)(\Sigma,\c_{pr},{\cal L})$ --- where
$U^*:\underline{JCLat}\to\underline{Res}_0$ is the functor that
together with
$U$ constitutes the equivalence of $\underline{JCLat}$ and
$\underline{Res}_0$ --- is the essentially unique canonical
resolution determined by $(\Sigma,\c_{pr},{\cal L})$. 
\ecl
\bpf   Surjectivity on objects implies that $V$ is
isomorphism-dense$^3$. A functor that is full, faithful and
isomorphism-dense describes the equivalence of its domain and
codomain [1,5,19].
\epf  By Theorem \ref{factorization}, an equivalent
characterization for
${\cal C}_{pr}:{\cal P}(\Sigma)\rightarrow{\cal L}$ is given by
$(\Sigma,{\cal L},{\cal C},\theta)$ such that
${\cal C}_{pr}=\theta\circ{\cal C}$. As such, the collection of
all operational resolutions gives rise to a bijective collection
of such quadruples. The following lemma shows that the morphisms
between operational resolutions can be characterized with the aid
of only the closure-part of the respective operational
resolutions.
\blm\label{propequivcondgeneral} For a map $f:{\cal
P}(\Sigma_1)\rightarrow{\cal P}(\Sigma_2)$ meeting
$A_{\cup}$ and two operational resolutions
${\cal C}_{pr,i}=\theta_i\circ{\cal C}_i:{\cal
P}(\Sigma_i)\rightarrow{\cal L}_i$ there is an equivalence of
condition
$A_{\#}$ with:
\par\medskip\noindent 
$A_*$: $\forall T\in{\cal P}(\Sigma_1): f({\cal
C}_1(T))\subseteq{\cal C}_2(f(T))$.
\elm
\bpf $A_{\#}$ implies $A_*$: \ ${\cal C}_{pr,1}(T)={\cal
C}_{pr,1}({\cal C}_1(T))\Rightarrow{\cal C}_{pr,2}(f(T))={\cal
C}_{pr,2}(f({\cal C}_1(T)))\Rightarrow{\cal C}_2(f(T))={\cal
C}_2(f({\cal C}_1(T)))\Rightarrow f({\cal C}_1(T))\subseteq{\cal
C}_2(f(T))$. Conversely, $A_*$ implies
$A_{\#}$: ${\cal C}_{pr,1}(T')={\cal C}_{pr,1}(T)\Rightarrow{\cal
C}_1(T')={\cal C}_1(T)\Rightarrow f({\cal C}_1(T'))=f({\cal
C}_1(T))\Rightarrow{\cal C}_2(f({\cal C}_1(T')))={\cal
C}_2(f({\cal C}_1(T)))\Rightarrow {\cal C}_2(f(T'))={\cal
C}_2(f(T))
\Rightarrow {\cal C}_{pr,2}(f(S))={\cal C}_{pr,2}(f(T))$.
\epf
\bpr\label{*****} We can define a quantaloid
$\underline{Res}_{\emptyset}^*$, in which joins of maps are
computed pointwise, by taking as object class the collection of
quadruples 
$(\Sigma,{\cal L},{\cal C},\theta)$ such that
${\cal C}_{pr}=\theta\circ{\cal C}:{\cal
P}(\Sigma)\rightarrow{\cal L}$ is an operational resolution, and
taking as hom-set between any two such objects
$(\Sigma_1,{\cal L}_1,{\cal C}_1,\theta_1)$ and $(\Sigma_2,{\cal
L}_2,{\cal C}_2,\theta_2)$:
$$\{f:{\cal P}(\Sigma_1)\rightarrow{\cal P}(\Sigma_2)\mid
im(f)=\emptyset 
\ or \ f \ meets \ A_{\cup},A_{\emptyset},A_{*}\}.$$ Further,
$\underline{Res}_{\emptyset}^{\#}$ and
$\underline{Res}_{\emptyset}^*$ are isomorphic categories.
\epr The following is obvious.
\bpr\label{V}  Setting that
$V:\underline{Res}_{\emptyset}^{*}
\rightarrow\underline{Clos}_{\emptyset}$
works on objects $(\Sigma,{\cal L},{\cal C},\theta)$ and 
$f\in \underline{Res}_{\emptyset}^{*}((\Sigma_1,{\cal L}_1,{\cal
C}_1,\theta_1),(\Sigma_2,{\cal L}_2,{\cal C}_2,\theta_2))$ as:
\[
\left\{\begin{array}{l} V(\Sigma,{\cal L},{\cal
C},\theta)=(\Sigma,{\cal C})\\ V(f):{\cal
P}(\Sigma_1)\rightarrow{\cal P}(\Sigma_2):T\mapsto f(T)
\end{array}
\right.  
\]
\noindent defines a fully faithful surjective quantaloid morphism.
\epr
\bpf 
$V(\Sigma,{\cal L},{\cal C},\theta)$ is a closure space for which
${\cal C}(\emptyset)=\emptyset$. Surjectivity on objects: for a
$\underline{Clos}_{\emptyset}$-object $(X,{\cal C})$ we have
evidently that
$V(X,{\cal F}(X),{\cal C}, id_{{\cal F}(X)})=(X,{\cal C})$, cfr.
Example
\ref{closuresasresolutions}.
$V$ is the ``identity'' on underlying maps of morphisms so
nothing to verify there.
\epf
\bcl\label{CL222}
$\underline{Res}_{\emptyset}^*$ and
$\underline{Clos}_{\emptyset}$ are categorically equivalent.
\ecl Using the material of the appendix, we can summarize this
categorical setting by the following scheme of quantaloids and
quantaloid morphisms:
\[
\begin{array}{ccc}  
\underline{Res}_{\emptyset}^{\#} &
\stackrel{iso}{\longleftrightarrow} &
\underline{Res}_{\emptyset}^{*}\\
\downarrow & & \ \ \ \updownarrow \cong \\  \underline{Res}_0 & &
\underline{Clos}_{\emptyset}\\
\cong\updownarrow\ \ \  & & \downarrow \\  \underline{JCLat}_0 &
\hookleftarrow &
\underline{Ints}y\underline{s}_{\emptyset}\\
\end{array}
\]
\par\vskip 0.406 truecm\par
\par\vskip 0.406 truecm\par
\noindent   {\bf 4. EXTENDING TO RESOLUTION OPERATORS}
\par\vskip 0.406 truecm\par
\par 
\noindent In the foregoing, it is apparent that the third
condition in the definition of `operational resolution' --- see
Eq.(\ref{df:opres3'}) --- has for consequences that:
\bit
\item the closure factor of an operational resolution, cfr.
Theorem
\ref{factorization}, is such that
${\cal C}({\emptyset})=\emptyset$;
\item the morphisms in the category
$\underline{Res}_{\emptyset}^{\#}$ must satisfy
$A_{\emptyset}$ in order to be structure preserving; likewise for
the morphisms in the category $\underline{Res}_{\emptyset}^*$;
likewise for the morphisms in the category
$\underline{Clos}_{\emptyset}$ to make Proposition \ref{V} work;
\item the morphisms in the category $\underline{Res}_0$ must
satisfy $A_0$ in order to make Proposition
\ref{Fpr} work; likewise for the morphisms in the category
$\underline{JCLat}_0$ to make Proposition  
\ref{U} work.  
\eit  Its motivation is primarily that of a conservation law,
expressing that whenever we have a physical system in some state
beforehand, we still have a physical system in some state after a
possible transition. However, one easily verifies that in our
constructions, we only need this ``empty kernel condition'' to
prove other ``empty kernel conditions''. In other words, we can
develop a completely analogous scheme without this condition,
giving rise to more general objects and more general morphisms. 
Referring to the previous section, the as such included
transitions with non-empty kernels can be interpreted as
``partially absurd transitions'' where only the subset of the
state space assuring a non-empty image is of physical relevance. 
Purely mathematical, this construction extends in this way the
morphismsets of the categories in [14,20] --- see   also our
appendix on this aspect.  
\bdf    A `resolution operator' from a set $\Sigma$ to a poclass
$({\cal L},\leq)$ is a map
$\r:{\cal P}(\Sigma)\rightarrow {\cal L}$ such that for all
$T,T',T_i\in{\cal P}(\Sigma)$:
\beqn T\subseteq T'\ \ \ \ \ \! \ \ \ \ \ \Rightarrow\ \ \ \ \
\r(T)\leq\r(T')\ \ \\
\forall i\in I:\r(T_i)\leq\r(T)\ \ \ \Rightarrow\ \ \, \ \
\r(\cup_i T_i)\leq\r(T)
\eeqn  
\edf The image of $\r$ is a complete join semilattice with
$\vee_i\r(T_i)=\r(\cup_iT_i)$, a generating set $\{\r(t)\mid
t\in\Sigma\}$, as bottom
$\r(\emptyset)$ and as top $\r(\Sigma)$. We have the following
result in analogy to Theorem \ref{factorization}.  
\bth  Any resolution operator $\r:{\cal P}(\Sigma)\rightarrow
{\cal L}$ factors uniquely into a closure operator
$\c:\p(\Sigma)\rightarrow\p(\Sigma)$ on $\Sigma$, and a
po-inclusion of the
$\c$-closed subsets $\f(\Sigma)$ into ${\cal L}$, say
$\theta:\f(\Sigma)\hookrightarrow {\cal L}$, such that
$\f(\Sigma)\cong im(\theta)$.
\eth Note that a resolution operator $\r$ is a $T_0$-resolution
($T_1$) if and only if the closure factor $\c$ is so.  For any
two given resolution operators
$\r_1=\theta_1\circ\c_1:\p(\Sigma_1)\rightarrow {\cal L}_1$ and
$\r_2=\theta_2\circ\c_2:\p(\Sigma_2)\rightarrow {\cal L}_2$, we
recall for a map 
$f:\p(\Sigma_1)\rightarrow\p(\Sigma_2)$:
\par\medskip\noindent
$A_{\cup}$: $f(\cup_iT_i)=\cup_if(T_i)$;
\par\noindent
$A_{\#}$: $\r_1(T)=\r_1(T')\Rightarrow\r_2(f(T))=\r_2(f(T'))$
\par\noindent
$A_{*}$: $\forall
T\subseteq\Sigma_1:f(\c_1(T))\subseteq\c_2(f(T))$
\par\medskip\noindent and for $g:im(\r_1)\rightarrow im(\r_2)$:
\par\medskip\noindent
$A_{\vee}$: $f(\vee_ia_i)=\vee_if(a_i)$.    
\par\medskip\noindent We have the following results in analogy to
Propositions \ref{propquantaloid},
\ref{Fpr} and
\ref{*****} and Corollaries \ref{CL111} and \ref{CL222}.
\bpr (i) We can define a quantaloid $\underline{Res}^{\#}$ with
as objects resolution operators, written as triples
$(\Sigma,{\cal L},\r)$, and with morphisms
$f:(\Sigma_1,{\cal L}_1,\r_1)\rightarrow(\Sigma_2,{\cal
L}_2,\r_2)$ determined by corresponding underlying maps
$f:\p(\Sigma_1)\rightarrow\p(\Sigma_2)$ that meet $A_{\cup}$ and
$A_{\#}$. The join of morphisms is computed pointwise.
\par\noindent (ii) We can define a quantaloid $\underline{Res}$
with as objects resolution operators, again written as triples
$(\Sigma,{\cal L},\r)$, and with morphisms
$g:(\Sigma_1,{\cal L}_1,\r_1)\rightarrow(\Sigma_2,{\cal
L}_2,\r_2)$ determined by corresponding underlying maps
$g:im(\r_1)\rightarrow im(\r_2)$ that meet
$A_{\vee}$. The join of morphisms is computed pointwise. Setting
for an object $(\Sigma,{\cal L},\r)$ and a morphism
$f:(\Sigma_1,{\cal L}_1,\r_1)\rightarrow(\Sigma_2,{\cal
L}_2,\r_2)$ of
$\underline{Res}^{\#}$ that:
\[
\left\{\begin{array}{l} F_{\r}(\Sigma,{\cal L},\r)=(\Sigma,{\cal
L},\r) \\ F_{\r}(f):im(\r_1)\rightarrow
im(\r_2):\r_1(T)\mapsto\r_2(f(T))
\end{array}
\right.
\] defines a full bijective quantaloid morphism
$F_{\r}:\underline{Res}^{\#}\rightarrow\underline{Res}$. Further,
$\underline{Res}$ is equivalent to $\underline{JCLat}$.
\par\noindent  (iii) We can define a quantaloid
$\underline{Res}^*$ with as objects resolution operators, now
written as quadruples $(\Sigma,{\cal L},\c,\theta)$, and with
morphisms
$f:(\Sigma_1,{\cal L}_1,\c_1,\theta_1)\rightarrow(\Sigma_2,{\cal
L}_2,\c_2,\theta_2)$ determined by corresponding underlying maps
$f:\p(\Sigma_1)\rightarrow\p(\Sigma_2)$ that meet $A_{\cup}$ and
$A_*$. The join of morphisms is computed pointwise. This
quantaloid is   isomorphic to $\underline{Res}^\#$ and equivalent
to $\underline{Clos}$.
\epr To summarize:
\[
\begin{array}{ccc}  
\underline{Res}^{\#}&
\stackrel{iso}{\longleftrightarrow} & \underline{Res}^{*}\\
\downarrow & & \ \ \ \updownarrow \cong \\  \underline{Res} & &
\underline{Clos} \\
\cong\updownarrow\ \ \  & & \downarrow \\  \underline{JCLat} &
\hookleftarrow &
\underline{Ints}y\underline{s}\\
\end{array}
\] which is exactly the same scheme that closed the previous
subsection, however without the ``empty kernel conditions''.
\par\vskip 0.406 truecm\par
\par\vskip 0.406 truecm\par
\noindent   {\bf 5. CONCLUSION: ON POSSIBLE STATE TRANSITIONS}    
\par\vskip 0.406 truecm\par
\par 
\noindent In the introduction we already sketched the reasoning
in [15] which assures that properties propagate with preservation
of the join.  In that same paper it is shown that with almost no
requirements it is possible to derive the unitary evolution of a
particle if one assumes strong determinism, i.e., when $f$ sends
states --- being the atoms of the supposedly complete atomistic
orthomodular property lattice --- onto states. However, this
hypothesis of ``strong deterministic evolution'' disables us to
express indeterministic transitions that do occur when
considering for example a perfect quantum measurement of the
property $a$ and its orthocomplement
$a'$, where the propagation of the `possible states' is described
by the following map:     
\[  f:{\cal P}(\Sigma)\to{\cal P}(\Sigma): 
\left\{
\begin{array}{l}
\{p\} \mapsto \{ a\wedge(a'\vee p) , a'\wedge(a\vee p)
\}\vspace{1mm}    
\\  T \mapsto \cup\{ f(\{p\}) \mid p \in T \}
\end{array}  
\right.    
\]  
\par
\noindent  provided that $T\cup\{0\}$ is interpreted as `possible
states' $T$ since $0$ is never true. This map sends any possible
initial state on its two possible outcome states $a\wedge(a'\vee
p)$ and $a'\wedge(a\vee p)$, formally expressed as
$f$ being the union of the maps that are atomically generated$^4$
by the respective Sasaki projectors. Clearly
$f$ cannot be ``reduced'' to a join preserving map between
property lattices, but one can verify that it does satisfy our
definition of a state transition   --- one can indeed prove that
any union of maps that are atomically generated by join
preserving maps satisfies $A_\#$ and $A_\cup$.  Since this
particular state transition occurs in standard quantum theory,
being an ordinary measurement described by a self-adjoint
operator with eigenspaces corresponding to $a$ and $a'$, it
clearly cannot suffice to work in a mathematical category where
the morphisms representing state transitions are join preserving
maps between the atomistic property lattices, as is implicitly
the case in [15,29].  In the case of our --- quantaloid ---
duality of categories, one category has the `physically
justifiable definite property transitions' --- described by join
preserving maps between property lattices --- as morphisms, and
the other category has the `underlying possible state
transitions' as morphisms, all this allowing the description of
indeterministic evolutions and as such generalizing the strong
deterministic evolutions to arbitrary ones. Within this context
we also mention an application related to linear logic [16,32],
which provides a syntactical tool to describe the above mentioned
perfect quantum measurement of $a$ and $a'$ [10].  More general,
the mathematical scheme presented in this paper delivers a class
of semantical interpretations for the corresponding logic that
describes the process of indeterministic propagation of states
for entities with a not necessarily Boolean description.  As a
present topic of further study we also mention the implications
of aspects of weak modularity and orthocomplementation within our
scheme, considering what already has been done for general
algebraic quantales [22,23,24].
\par\vskip 0.406 truecm\par
\par\vskip 0.406 truecm\par
\noindent   {\bf 6. DISCUSSION: DESCRIBING COMPOUNDNESS}
\par\vskip 0.406 truecm\par
\par 
\noindent  In this section we discuss the use of the morphismsets
in our dual quantaloids for the description of compound systems
in the spirit of [6,7,8].   In particular, with the hypothesis
that any kind of interaction between two systems boils down to
the fact that `actuality of a property $a_1$ of the first system
is caused by to actuality of property $a_2$ of the second
system', we can recover the rays of the tensor product of Hilbert
spaces as the description of the `states of compoundness' for
corresponding quantum systems [9].    It can indeed again be
argued that also in this case, the corresponding maps ---
describing mutual induction of properties --- should be join
preserving: with any two interacting physical systems,
respectively described by lattices of  verifiable properties
${\cal L}_1$ and ${\cal L}_2$, we can associate a map
$f^*: {\cal L}_2
\to {\cal L}_1$ with
$a_1=f^*(a_2)$ being the cause of $a_2$ in ${\cal L}_1$, that is,
$a_1$ is the weakest property in
${\cal L}_1$ whose actuality causes the actuality of $a_2$; all
this again assures a join preserving Galois dual $f$ that
expresses mutual induction of properties of one system onto the
other.  Then again, by applying the same tools as in [15], i.e.,
the general theory of morphisms of projective geometries [14] in
combination with Piron's representation theorem [25,26], it is
possible to prove that one obtains a complete lattice, with the
anti-Hilbert-Schmidt maps as atoms --- the atoms of the obtained
complete lattice are exactly the join preserving maps that send
atoms on atoms or the bottom element --- and
$[{\cal L}_1\setminus\{0_1\}\to {\cal L}_2:a_1\mapsto 1_2;
0_1\mapsto 0_2]$ as top element [9]:    
\par
\medskip 
\noindent         (i) We can interpret the anti-Hilbert-Schmidt
maps between Hilbert space
$H_1$ and $H_2$ as `states of maximal compoundness' since they
correspond in a one to one way with the rays in $H_1\otimes
H_2$.          
\par
\medskip 
\noindent   (ii)  The top element can be interpreted as the
`state of separation' --- differently discussed in [2], where
separation refers to a type of entity and not to a state of a
compound system --- since it expresses that actuality of a
property of the first system implies ``existence'' of the second
and --- strictly --- nothing more, this assuming that both
systems exist {\it a priori}.  
\par
\medskip 
\noindent As such, the morphismsets in our quantaloids generalize
the description of the interaction between individual entities
within compound systems.    One can proceed the same reasoning
for general compound systems consisting   of any number of
individual entities.  It follows that a general description for
compound systems corresponds with a commuting diagram within the
two dual quantaloids: the arrows in the diagram represent the
mutual induction of states and properties and the commutativity
follows from the requirement that there should be `structural
independence' on the order of performance of the measurements on
the individual entities within the compound system.  A paper on
the matter is in the publication pipeline.
\par\vskip 0.406 truecm\par
\par\vskip 0.406 truecm\par
\noindent   {\bf APPENDIX: QUANTALOIDS}
\par\vskip 0.406 truecm\par
\par 
\noindent  Below we give some mathematical preliminaries to the
content of this paper related to quantaloids. References are
[1,5,19] for categories and [18,28,31] for quantaloids.
\bdf\label{df:quantaloid}  A quantaloid is a category such that: 
\par\noindent (i) every hom-set is a join complete semilattice;
\par\noindent (ii) composition of morphisms distributes on both
sides over joins.
\par\noindent  Let $\underline{Q}$ and $\underline{R}$ be
quantaloids. A quantaloid morphism from $\underline{Q}$ to
$\underline{R}$ is a functor
$F:\underline{Q}\rightarrow\underline{R}$ such that on hom-sets
it induces join-preserving maps
$\underline{Q}(A,B)\rightarrow\underline{R}(FA,FB)$.
\edf In the language of enriched category theory [5] we can say
that a quantaloid is a category that is enriched in
$\underline{JCLat}$, the category of join complete semilattices
and join-preserving maps, and a quantaloid morphism is a
$\underline{JCLat}$-enriched functor. A quantaloid with one
object is commonly known as a `unital quantale' [22,30]. Another
point of view is that in a quantaloid every hom-set of
endomorphisms on an object (a hom-set of ``loops'') is a unital
quantale. The restriction of a quantaloid morphism to a hom-set
of 'loops' yields what is known as a `unital quantale morphism'.
The quantaloids that are constructed in this paper, are exactly
in this way generalizations of the unital quantales that are
constructed, and motivated physically, in [3,11]. 
\bex\label{jclatcat}  The category of join complete semilattices
and join-preserving maps $\underline{JCLat}$ is a quantaloid,
with respect to pointwise ordering of maps $[28]$.
\eex  As can easily be verified, any subcategory of a quantaloid
that is closed under the inherited join of morphisms, is a
subquantaloid. Thus any full subcategory of a quantaloid is a
subquantaloid, and selecting from a given a quantaloid certain
morphisms but keeping all the objects, gives rise to a
subquantaloid if and only if the inherited join of morphisms is
internal. Often, such a subquantaloid is constructed by imposing
extra conditions on the morphisms, verifying that these extra
conditions ``respect'' arbitrary joins.
\bex  Selecting from $\underline{JCLat}$ those morphisms
$f:{\cal L}\rightarrow{\cal M}$ that meet the extra condition
$f(a)=0_{\cal M}\Leftrightarrow a=0_{\cal L}$, we obtain a new
quantaloid since any join of such maps is again such a map. We
will denote this new quantaloid by
$\underline{JCLat}_0$.
\eex
\bex  Consider a category with as objects closure spaces
$(X,{\cal C})$, in which a morphism $f:(X_1,{\cal
C}_1)\rightarrow(X_2,{\cal C}_2)$ is represented by an underlying
union-preserving map between the respective powersets, that is,
$f:{\cal P}(X_1)\rightarrow{\cal P}(X_2)$ such that
$f(\cup_iT_i)=\cup_if(T_i)$. This is in fact a quantaloid in
which the join of maps is computed pointwise. Keeping all the
objects and selecting those morphisms that satisfy $\forall
T\in{\cal P}(X_1):f({\cal C}_1(T))\subseteq{\cal C}_2(f(T))$, we
obtain a subquataloid that we will denote by $\underline{Clos}$,
since the condition respects the join of morphisms. Now selecting
those closure spaces
$(X,{\cal C})$ for which ${\cal C}(\emptyset)=\emptyset$ and
those morphisms that send
$\emptyset$ exactly on $\emptyset$, that is,
$f(T)=\emptyset\Leftrightarrow T=\emptyset$, we obtain a
subquantaloid
$\underline{Clos}_{\emptyset}$ of $\underline{Clos}$, since the
extra condition on morphisms respects joins.
\eex  Another category of closure spaces that plays an important
role in for instance [14,15,20] has as objects all closure spaces
$(X,{\cal C})$ and as morphisms between
$(X_1,{\cal C}_1)$ and $(X_2,{\cal C}_2)$ all of the `continuous'
maps
$f:X_1\setminus K\rightarrow X_2$ defined on the complement of
$K\subseteq X_1$, that is
$f({\cal C}_1(T)\setminus K)\subseteq{\cal C}_2(f(T\setminus K))$
for all
$T\subseteq X$. Denoting this category as
$\underline{S}p\underline{ace}$, it is easy to see that there is
a functor 
$Ext:\underline{S}p\underline{ace}\rightarrow\underline{Clos}$
that is the identity on objects and:
\beqa Ext(f):{\cal P}(X_1)\rightarrow{\cal P}(X_2):
T\mapsto\{f(x)|x\in T\setminus K\}
\eeqa for a morphism
$f:(X_1,{\cal C}_1)\rightarrow(X_2,{\cal C}_2)$. But the morphisms
$Ext(f)$ meet the extra condition that
$Ext(f)(\{x\})$ is a singleton or the empty set for all $x\in X$.
Since this condition is not preserved by joins,
$\underline{S}p\underline{ace}$ is not a quantaloid and
$Ext$ is not a quantaloid morphism. However,
$\underline{S}p\underline{ace}$ can be embedded in
$\underline{Clos}$ and this embedding restricts to an embedding
$\underline{S}p\underline{ace}_{\emptyset}
\to\underline{Clos}_{\emptyset}$,
where
$\underline{S}p\underline{ace}_{\emptyset}$ is the category with
those objects of
$\underline{S}p\underline{ace}$ such that ${\cal
C}(\emptyset)=\emptyset$ and of which all morphisms have an empty
kernel. The same sort of remark can be made on the categories of
lattices. The typical category of lattices that one finds in
[14,15,20] is
$\underline{JCALat}$: its objects are complete atomistic
lattices, its morphisms are join complete lattices that send
atoms onto atoms or onto the bottom (the full subcategory
$\underline{T}_1\underline{S}p\underline{ace}$ of
$\underline{S}p\underline{ace}$ consisting of $T_1$-closures is
then equivalent to $\underline{JCALat}$ --- which is exactly the
core of the mathematical developments in both  [14,20]).
$\underline{JCALat}$ can be embedded into $\underline{JCLat}$ as
category, simply by a ``forgetful functor''
$U:\underline{JCALat}\to\underline{JCLat}$, but again it is
evidently not true that the join of morphisms
$U(f_i)$ is a morphism $U(f)$, because such a join does not
necessarily send atoms onto atoms. Of course, this embedding
restricts to an embedding
$U:\underline{JCALat}_0\to\underline{JCLat}_0$, where now 
$\underline{JCALat}_0$ is the subcategory of $\underline{JCALat}$
of  which  the morphisms never send an atom to the bottom.
\bex\label{Intsys}  By an `intersection system' we mean a
collection of subsets of a certain set $X$, ordered by inclusion,
closed under intersection. Any intersection system is a complete
lattice, ordered by     set-inclusion, thus we can define a
quantaloid
$\underline{Ints}y\underline{s}$ as the full subcategory of
$\underline{JCLat}$ of which the objects are intersection systems.
Accordingly we construct
$\underline{Ints}y\underline{s}_{\emptyset}$ as the full
subcategory of
$\underline{JCLat}_0$ of which the objects are intersection
systems with bottom element $\emptyset$.
\eex  
\bex  Setting that
$W:\underline{Clos}\rightarrow\underline{Ints}y\underline{s}$
works on
$(\Sigma,{\cal C})$ and
$f\in \underline{Clos}((\Sigma_1,{\cal C}_1),(\Sigma_2,{\cal
C}_2))$ respectively as:
\[
\left
\{\begin{array}{l} W(\Sigma,{\cal C})={\cal F}(\Sigma)\\
W(f):{\cal F}_1(\Sigma_1)\rightarrow{\cal F}_2(\Sigma_2):F\mapsto
{\cal C}_2(f(F))  
\end{array}  
\right.
\]
\noindent we have defined a full bijective quantaloid morphism.
The same is true for the obvious sub-functor
$W:\underline{Clos}_{\emptyset}\rightarrow
\underline{Ints}y\underline{s}_{\emptyset}$
\eex
\bpf Since a closure operator ${\cal C}$ on a set $\Sigma$ can be
characterized completely by the intersection system ${\cal
F}(\Sigma)$ of  
${\cal C}$-closed subsets, the bijectivity is clear. To show that
the action of $W$ is functoral and full requires some
verifications that are analogous to those of the proof of
Proposition \ref{Fpr}.
\epf
\par\vskip 0.406 truecm\par
\par\vskip 0.406 truecm\par    
\noindent   {\bf ACKNOWLEDGMENTS}  
\par\vskip 0.406 truecm\par  
\par 
\noindent  We thank D.J. Moore, C. Piron and F. Valckenborgh for
discussions related to the content of this paper and Cl.-A. Faure
and J. Paseka for reading it and adding interesting comments. I.
Stubbe thanks D. Aerts and FUND-DWIS-VUB for logistic and
financial support during the realization of the results in this
paper. B. Coecke is Post-Doctoral Researcher at Flanders' Fund
for Scientific Research.    
\par\vskip 0.406 truecm\par
\par\vskip 0.406 truecm\par
\noindent   {\bf REFERENCES}\footnote{Papers by the present authors are downloadable
at (.ps files with fonts): http://www.vub.ac.be/CLEA/BobDownloads.html}  
\par\vskip 0.406 truecm\par
\par 
\noindent  1.\hspace{2mm}J$.$ Adamek, H$.$ Herrlich and G.E$.$
Strecker, {\it Abstract and Concrete Categories}, John Wiley \&
Sons (1990).
\par\vskip 0.1 truecm\par 
\noindent  2.\hspace{2mm}D$.$ Aerts, {\it Found$.$ Phys$.$} {\bf
24}, 1227 (1982).
\par\vskip 0.1 truecm\par 
\noindent  3.\hspace{2mm}H$.$ Amira, B$.$ Coecke and I$.$ Stubbe,
{\it Helvetica Phys$.$ Acta} {\bf 71}, 554 (1998).
\par\vskip 0.1 truecm\par   
\noindent  4.\hspace{2mm}G$.$ Birkhoff, {\it Lattice Theory}, AMS
Coll$.$ Publ$.$ (1940).
\par\vskip 0.1 truecm\par 
\noindent  5.\hspace{2mm}F$.$ Borceux, {\it Handbook of
Categorical Algebra Part 1 and 2}, Cambridge University Press
(1994). 
\par\vskip 0.1 truecm\par 
\noindent  6.\hspace{2mm}B$.$ Coecke, {\it Helvetica Phys$.$
Acta} {\bf 68}, 394 (1995).
\par\vskip 0.1 truecm\par 
\noindent  7.\hspace{2mm}B$.$ Coecke, {\it Found$.$ Phys$.$} {\bf
28}, 1109 (1998).
\par\vskip 0.1 truecm\par 
\noindent  8.\hspace{2mm}B$.$ Coecke, {\it Found$.$ Phys$.$} {\bf
28}, 1347 (1998).
\par\vskip 0.1 truecm\par 
\noindent  9.\hspace{2mm}B$.$ Coecke, ``Structural
Characterization of Compoundness'', {\it Int$.$ J$.$
Theor$.$ Phys.} {\bf 39}, 585 (2000).
\par\vskip 0.1 truecm\par 
\noindent  10.\hspace{2mm}B$.$ Coecke and S$.$ Smets, ``A Logical
Description for Perfect Measurements'', {\it Int$.$
J$.$ Theor$.$ Phys.} {\bf 39}, 595 (2000).
\par\vskip 0.1 truecm\par 
\noindent  11.\hspace{2mm}B$.$ Coecke and I$.$ Stubbe, ``On a
Duality of Quantales emerging from an Operational Resolution'',
{\it Int$.$ J$.$ Theor$.$ Phys.} {\bf 38}, 3269 (1999).
\par\vskip 0.1 truecm\par 
\noindent  12.\hspace{2mm}W$.$ Daniel, {\it Helvetica Phys$.$
Acta} {\bf 62}, 941 (1989).
\par\vskip 0.1 truecm\par 
\noindent  13.\hspace{2mm}D.J$.$ Foulis, {\it Proc$.$ AMS} {\bf
11}, 648 (1960).  
\par\vskip 0.1 truecm\par 
\noindent  14.\hspace{2mm}Cl.-A$.$ Faure and A$.$ Fr\"olicher,
{\it Geom. Dedicata} {\bf 47}, 25 (1993).
\par\vskip 0.1 truecm\par 
\noindent  15.\hspace{2mm}Cl.-A$.$ Faure, D.J$.$ Moore and C$.$
Piron, {\it Helvetica Phys$.$ Acta} {\bf 68}, 150 (1995).
\par\vskip 0.1 truecm\par 
\noindent  16.\hspace{2mm}J.Y$.$ Girard, {\it Theor$.$ Comp$.$
Sc$.$} {\bf 50} (1987).
\par\vskip 0.1 truecm\par 
\noindent  17.\hspace{2mm}P.T$.$ Johnstone, {\it Stone Spaces},
Cambridge University Press (1982).
\par\vskip 0.1 truecm\par 
\noindent  18.\hspace{2mm}A$.$ Joyal and M$.$ Tierney, {\it Mem.
AMS} {\bf 51}, No.309 (1984).
\par\vskip 0.1 truecm\par 
\noindent  19.\hspace{2mm}S$.$ MacLane, {\it Categories for the
Working Mathematician}, Springer-Verlag (1971/1997).
\par\vskip 0.1 truecm\par 
\noindent  20.\hspace{2mm}D.J$.$ Moore, {\it Helvetica Phys$.$
Acta} {\bf 68}, 658 (1995).
\par\vskip 0.1 truecm\par 
\noindent  21.\hspace{2mm}D.J$.$ Moore, {\it Int$.$ J$.$ Theor$.$
Phys.} {\bf 36}, 2211 (1997).
\par\vskip 0.1 truecm\par 
\noindent  22.\hspace{2mm}C.J$.$ Mulvey, {\it Rend$.$ Circ$.$
Math$.$ Palermo} {\bf 12}, 99 (1986).
\par\vskip 0.1 truecm\par 
\noindent  23.\hspace{2mm}C.J$.$ Mulvey and J$.$ Wick-Pelletier,
{\it Canadian Math$.$ Soc$.$ Conf$.$ Proc$.$} {\bf 13}, 345
(1992). 
\par\vskip 0.1 truecm\par 
\noindent      24.\hspace{2mm}J$.$ Paseka, ``Simple Quantales'',
ed$.$ P$.$ Simon, {\it Proceedings of the 8th Prague Topology
Symposium}, p.314 (1996).
\par\vskip 0.1 truecm\par 
\noindent  25.\hspace{2mm}C$.$ Piron, {\it Helvetica Phys$.$
Acta} {\bf 37}, 439 (1964).
\par\vskip 0.1 truecm\par 
\noindent  26.\hspace{2mm}C$.$ Piron, {\it Foundations of Quantum
Physics}, W.A$.$ Benjamin (1976).  
\par\vskip 0.1 truecm\par 
\noindent  27.\hspace{2mm}C$.$ Piron, {\it J$.$ Phyl$.$ Logic}
{\bf 6}, 481 (1977).
\par\vskip 0.1 truecm\par 
\noindent  28.\hspace{2mm}A.M$.$ Pitts, {\it Proc$.$ London
Math$.$ Soc$.$} {\bf 57}, 433 (1988).
\par\vskip 0.1 truecm\par 
\noindent  29.\hspace{2mm}J.C.T$.$ Pool, {\it Comm$.$ Math$.$
Phys$.$} {\bf 9}, 118 (1968).
\par\vskip 0.1 truecm\par 
\noindent  30.\hspace{2mm}K.I$.$ Rosenthal, {\it Quantales and
Their Applications}, Pitmann Research Notes in Math$.$ {\bf 234},
{\it Longmann Sc$.$ \& Tech$.$ Publ$.$} (1990).
\par\vskip 0.1 truecm\par 
\noindent  31.\hspace{2mm}K.I$.$ Rosenthal,  {\it J$.$ Pure
Appl$.$ Alg$.$} {\bf 77}, 67 (1991).    
\par\vskip 0.1 truecm\par 
\noindent    32.\hspace{2mm}D.N$.$ Yetter, {\it J$.$ Symb$.$
Logic} {\bf 55}, 41 (1990).
\par\vskip 0.406 truecm\par  
\par\vskip 0.406 truecm\par    
\noindent   {\bf NOTES}
\par\vskip 0.406 truecm\par
\par 
\noindent 1. This formulation --- discussed with D.J. Moore 
privately --- differs from the one in [15] the sense that it does
not make any reference to the tests that define properties in an
operational way [2,15,20,25,26].
  
\par\medskip\noindent 2. 'Poclass' is short for 'partially
ordered class', being a thin category
${\cal L}$ in which any two different objects are non-isomorphic,
wherein we   write $a\leq b$ if and only if there is (exactly)
one morphism from
$a$ to $b$. Since Definition \ref{opres} makes no reference to
the whole of  
${\cal L}$ but only to at most set-many elements of
${\cal L}$, we can indeed work with a poclass rather than a poset
for the codomain ${\cal L}$. The partial ordering on the codomain
${\cal L}$ can be operationally motivated [26].  The fact that
${\cal L}$ might be larger than
$im({\cal C}_{pr})$ is essential: we need to be able to consider
one
${\cal L}$ for different
$\Sigma$'s and
${\cal C}_{pr}$'s, with not coinciding images, allowing the joint
consideration of the properties of a compound system and those of
its subsystem. 

\par\medskip\noindent 3. A functor
$F:\underline{A}\rightarrow\underline{B}$ is called
isomorphism-dense if for any $\underline{B}$-object $B$ there
exists some
$\underline{A}$-object $A$ such that $F(A)\cong B$.

\par\medskip\noindent 4. A map is atomically generated when the
image of
$T\subseteq\Sigma$ is the union of the images of
$p\in\Sigma$ by the underlying atomic map, in this case a Sasaki
projection. For details we refer to [3].
  
\end{document}